\documentclass[journal]{IEEEtran}
\usepackage{cite}
\usepackage{amsmath,amssymb,amsfonts}
\usepackage{balance} 
\usepackage{graphicx}
\usepackage{algorithmic}
\usepackage{algorithm}
\usepackage{amsthm}
\usepackage{pslatex}			
\usepackage{latexsym}
\usepackage{subfig}
\usepackage{textcomp}
\usepackage{xcolor}
\usepackage{float}
\usepackage[compact]{titlesec}
\ifCLASSINFOpdf
\else
\fi
\begin{document}
%
\title{Image Based Password Authentication System}
%
%
%

\author{Sanjida Akter Sharna$^{1, 2, 3}$, Sheikh Ashraf Ali$^{1, 2, 4}$\\
\normalsize $^1$Military Institute of Science and Technology, Dhaka-1000, Bangladesh\\
\normalsize $^2$Department of CSE,Military Institute of Science and Technology\\

Email: $^3$sharnasanjida@gmail.com, $^4$ashrafalimugdho@gmail.com\\
}

\maketitle

\begin{abstract}
Preservation of information and computer security is broadly
dependent on the secured authentication system which is underpinned
by password. Text based password is a commonly used and available
system for authentication. But it bears many limitations like shoulder
surfing, dictionary attack, Phishing, guessing the password etc. In order to overwhelm these vulnerabilities of ancient textual password, many graphical or image based password authentication system
has been introduced form last few years. But none of this graphical system is considered as enough adventurous to keep pace with these issues.
Here we have proposed an image based password authentication system
which is more methodical and can cope up with every vulnerability of
recent password authentication system. To make our system hassle free
and more reliable, we will only take username from an user for registration purpose as our system will generate a unique key number for that
particular user and this key will be used as password for later login procedure.
The user name and key both will be encrypted using a cryptography algorithm to
prevent database hacking. There will be a randomized clickable image grid in our system. By clicking on this image grid, user will input the password key for login purpose. Here we have developed another method namely ‘shoulder surfing resistant password’. To prevent the attack of shoulder surfing, if any user wishes to change our system provided password key then he/she is allowed to do so by using this method. Besides this method allows user to change the password every single time of login. A user doesn't need to enter any textual password for authentication in our recent module and hence combination of all these features improve the security, usability and user friendliness of our system.

\end{abstract}

\begin{IEEEkeywords}
Image, Password, Authentication, Shoulder surfing, Graphical.
\end{IEEEkeywords}

%
\IEEEpeerreviewmaketitle

\section{Introduction}
%
%
%
%
\IEEEPARstart{A}{uthentication} system is of great importance from the raise of information and
technology for the confidentiality of data, information, statistics and many other
stuffs of individuals or any organizations. From the ancient era, textual based
authentication system has been used for the security purpose. But now with
the proliferation of technology and advancement, hackers have become smart
enough to decrypt any type of textual password. They have found many ways
like password guessing, dictionary attack, hash guessing, rainbow table, password
sniffing, brute force attack and many other efficient ways to crack any strong
password\cite{b11}. Considering all these reasons, the information and data under this type
of faulty authentication system are at great risk of disclosing\cite{b12}. To get rid of this
problem, graphical authentication system has been introduced a few years ago.
There are many types and categories of graphical authentication systems which
have been launched to preserve confidentiality of information. But none of those
systems are enough systematic to protect data fully. Now a days shoulder surfing
is the main obstacle to this graphical authentication system. If any observer
observes the graphical login system for some time then he could probably guess
the pattern or type of graphical password which leads the main failure of this
authentication system\cite{b8}. Recalling all these impediments, we have designed such a
system which is shoulder surfing registrant and not textual. During registration
process, our system will provide system generated a key number to user which
will be required for login as password. While login, user has to provide the user name and the key number which will be taken using special technique. That is the user has to click on the images in a 10x10 image grid to input key number. The image grid is used to input pass key number from user, so that anyone is spying on the user can not get the key number. The user can also make his own
key number by adding the current date to the previous key number which is named as shoulder surfing registrant password. Throughout
the authentication system, the main objective is to develop such a module which
is not fully textual so that no hackers can easily crack the password using
the latest technique of password decryption. Another prime objective is that to
get rid of the failure (shoulder surfing) of graphical authentication system\cite{b7}.

\section{Literature review}

\subsection{A Shoulder surfing resistant graphical password scheme-WIW}
Recently a shoulder resistant scheme was introduced where they describe their
system as follows: In this system, they mainly used four factors. A user needs to
choose ‘h’ number of images to make his password. This variable is dependent on
each user. They can choose it according to their wish. While log-in the content of
this h images will change randomly. These images are described as scenes here.
For each scene, there will be ‘n’ objects which have fix ranging from 250-300.
Then k is defined as pass-object which is only chosen by the user and it is the
part of the password. Then another factor m is defined as perturbation i,e the
appearance of each pass object in each scene. While log-in the user must be
able to recognize those pass objects. One image is stand for one letter, which
is determined by the system. Now user need to identify the appearances and
locations of pass objects in each scene. This combination spells a letter for each
pass object. This way user needs to identify all the combination which spells
some letter. The the main task is to identify all the letters for successful log-in.
Each time of log-in, this pass objects and non pass objects and their appearances
will be randomized. For the attacker, it is difficult to identify those letters and
recognize the pattern as each time it is randomized. The main problem with this
system is that, if the value of h, k and m is smaller than its quite easy for a user
to memorize those factors but then the attackers may recognize the pattern while
studying for some days. So it weakens the system. If the user chooses large value
of h,k and m, then for log-in the user needs to memorize a lots of factors like
all the pass objects form each scene accurately. In case, the user fails to choose
the pass objects correctly then log-in process will be unsuccessful. So here the
whole system is not at all user friendly or not usable for ordinary users \cite{b1}.
In our system the user has to face no difficulties to form his password, he
just needs to provide his user name which is comparatively easy to memorize.
For the user, our system will generate a key number and provide him. As it is
generated by the system,it will be highly secured. Then while log-in the user
just need to enter the key number from image grid.

\subsection{A shoulder surfing resistant image based authentication system
with temporal indirect image selection}

In this system, it displays images in sequence like slide show and an user needs
to select slide show which includes his pass image among several slides show.
The system presents several images to user. user selects several pass images p,
which is altogether used as his password. And the user must remember this pass
images. In log-in phase the system present N slide shows and M is the number
of images in each slide show. Now the user needs to recognize in which slide
show the pass image is included, then he needs to select the exact pass images
from each slide show which changes with a fix interval of time 1/v per image.
By choosing the correct sequence, the log-in will be successful.
Now this system has almost 50 percentage threats of shoulder surfing. If any
hacker or attacker take a video clip of any user through CCTV camera or somehow, then after studying for some days he can easily recognize the pattern.1st the correct slide show is selected then the pass image form that particular images
sets of slide show is selected. then by observing patiently, the attacker will be
aware of the pass images. Here the switching interval speed v is a great factor
which is in millisecond. If the switching interval is less or can not be controlled
properly then the threat of revealing the pass images is higher \cite{b2}.
\\After studying this paper, we observe that our system uses no such slide
shows or here switching interval or speed is not a factor, we are randomizing
our image grid which it is free of threat of revealing pass images. In our system,
we are using another additional method of entering key which is described as
below. User has to add the decimal digits of the log-in date in such a way that it
will form a single digit (0-9). By repeating this single digit 4 times it will form
a 4-digit number. then user has to add this repeated 4-digit number with the
original 4 digit key. If the summation is greater than10 then only last digit will
be taken. So, our system is highly secured of shoulder surfing problem.

\subsection{A review on two level authentication using image selection and
voice recognition}

In this system image and voice is used for authentication. During registration
user has to select an image and enter a watermark. This image with the watermark is one level of authentication. Then user needs to give the voice summon
which is the 2nd level of authentication. In this system, they used voice input
which is not very much user friendly. Different voice may occur during recognition \cite{b3}.
\\Our system is very much user friendly. We just need to remember a 4-digit
code to log in. In this system there is a high possibility of shoulder surfing while
selecting the image. User need to select just one image to log-in. So, it is easy for
attacker to know the required image by shoulder surfing. Our system is highly
preventable form shoulder surfing problem. We used a random image grid and
a shoulder surfing resistant key which is described previously to prevent this
problem.
\subsection{Image based authentication system}

In this system, both password and image are used to overcome the problems. Like
other authentication system, it also consists of two phases that is registration
and log-in. In registration phase user needs to fulfill basic personal requirements
like name, DOB, e-mail address etc. Next user needs to enter a password which
must maintain their constraint like minimum 8-character password, minimum
one uppercase and one numerical number and one special character. By full
filling these criteria, user can choose their desired password. After this stage
user must select one image as his pass image otherwise the registration will
be unsuccessful. There is an image category selection option, from where user
needs to select his desired category like natural scenarios, animals, flowers etc.
Every time these categories will be randomized by refreshing the page. Now if
user selects 3 images, then he/she needs to select the category, then from his
chosen category a 3*3 image grid of same type but discrete images will appear.
From this grid user will choose some of his pass images. Then he has to choose
his 2nd category. By choosing 2nd category, again a 3*3 grid of this category
will appear. Once again user needs to choose pass image from this category. In
the same manner user will choose the 3rd category and pass images. User must
memorize this pass images. In login phase, user must provide username, password
and pass images. Pass images will be randomly displayed in login round. In each
login round, there may be all pass images or some images. It is also possible that
none of pass images appear in any round. User has to choose the pass images
as the exact sequence he chooses them through registration phase. By choosing
the correct sequence from login rounds, login will be successful \cite{b4}.
\\In their system, they cannot prevent offline dictionary attack and the system is slower comparing traditional textual authentication system for displaying
image grid several time. They do not use any encryption system to encrypt the
password. If any hacker hacks the database then he can easily get the password
and user name and it can make one step easier for the hacker to hack the system.
And in their system, shoulder surfing is prevented but not in a wholesome manner.
\\Comparing with this system, our system almost prevents offline dictionary
attack and database hacking. As in our system the user name and password
which is a key number (generated by our system) will be encrypted through
AES and will be saved in database. So, if hacker somehow manages to hack the
database, he won’t get the exact password. As the hacker don’t know the key, he
won’t be able to decrypt it. Our image grid appears only once, and randomizes in
every click. For this reason, it is faster than the mentioned system. And because
of our advanced and randomized image grid, the possibilities of shoulder surfing
are almost prevented which is explained broadly below.
\subsection{A secure graphical password authentication system}

This system allows a user to create a graphical password by selecting an image
from a collection of available pictures. From the selected images user has to
select one grid as the password. The selected image is watermarked with a cover
image using Generic Visible Watermark Embedding technique. The method is
based on the use of deterministic one-to-one compound mappings of image pixel
values for overlaying a variety of visible watermarks of arbitrary sizes on cover
images. During login, after entering the user details a QR Code is generated
in the computer. User has to scan the QR code using his mobile phone. After
scanning, a collection of images will be appeared in the screen of the phone. User
has to select the image. After choosing correct image, the watermarked image
will be appeared on the screen. User has to choose the correct grid position that
he has already registered in the watermarked image \cite{b5}.
\\This system is not user friendly at all. Normal user won’t feel comfortable to
use such kind of inconvenient system. So, this system will be failed initially for
lacking of user friendliness. On the other hand, our system is quite user friendly.

\section{Our proposed system}
\subsection{Registration system architecture}
\subsubsection{Provide User Name}
 The initial phase of our authentication system is registration. In this phase like other authentication system, user needs to provide his
username which must be alphanumeric. Here we are providing an extra benefit
that is an user does not require to provide or create any password by thyself.
As the user doesn’t need to remind any password, our system will be more
comfortable and user friendly.
\subsubsection{Key number}
After providing user name, our system will generate a unique
key number which consists of 4 digit decimal number. User just has to memorize
this key number for login. In another word, we can say this key number can be
regarded as his password.

\subsubsection{AES (Advanced Encryption Standard) encryption}
The generated key and user name will be encrypted using the cryptography method AES-128\cite{b14}. After that the encrypted key and user name both will be saved in database so that if any hacker hacks the database, he will not get the actual user name and key number. As the hacker doesn’t know the AES key, he cannot easily decrypt the user name and key number.

\subsection{Login system architecture}
\subsubsection{Enter User Name}
If a user wants to login, then the system will initially ask for
his user name. A window will appear where user has to enter the accurate user name. If the user name is similar to his previously provided user name while-registration phase, then the system will let him to proceed to the next step. In
next step a 10x10 image grid will appear.

\subsubsection{10x10 image grid}
In this step, a 10x10 image grid\cite{b10} will appear. User must
enter the previously provided key number by clicking on this image grid\cite{b6}. Images of this grid are clickable except the images of 1st row. The image grid contains 25 original images and 3 copies of each images in total 100 images which will be arranged in 10 rows and 10 columns\cite{b9}. In the 1st row of the grid, 10 different images will be appeared randomly from the original 25 images. The 1st row represents a value from left to right according to its index number which is numbered as 0 to 9. The images which will appear on 1st row, their value will be set as index number. The duplicates of these 10 images will represent the same value of original images in the 1st row. The rest of the 15 original images and its copies represent garbage value. 
\begin{figure}[h]
\centering
\centerline{\includegraphics[width=8cm,height=7cm]{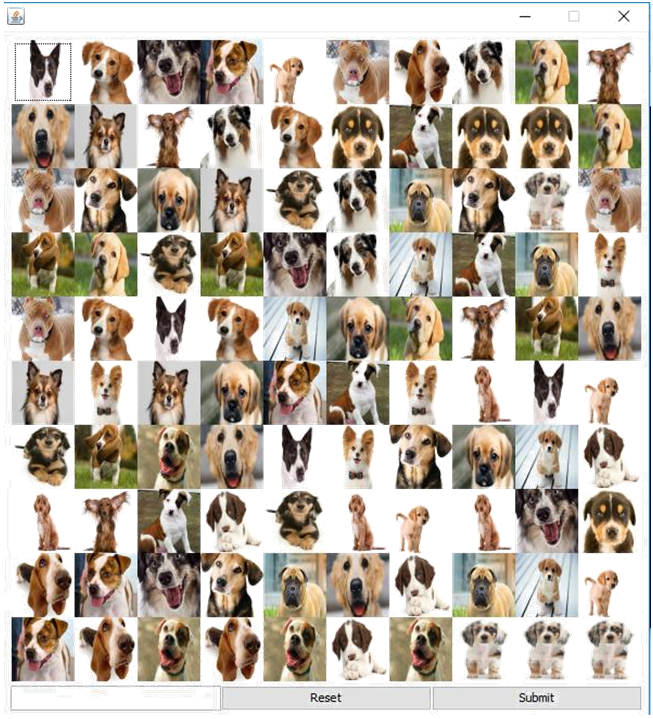}}
\caption{10x10 Image Grid} 
\label{fig1}
\end{figure}
For example if the key number is 6431, then the user needs to watch out the 6th index image of 1st row and then find the same image anywhere on the grid and click. The system will take this as input digit 6. Following the same method rest of the digits will be entered by clicking on the image grid. The images and 1st row of the image grid will randomize on every login session. When the user will click images to enter the 4 digit key number, by each click the whole image grid including the 1st row will be shuffled randomly so that any person or observer who is trying to assume the key number gets puzzled to find out the image pattern. As the grid will shuffle in each click, the images will change randomly every time. In case if the observer somehow possesses an eidetic memory/photographic memory, it will be very tough for him to recall every images correctly and he/she will be perplexed to find the relation of key number and image pattern. Hence hacker will be failed to guess the key number.

\subsubsection{Shoulder surfing resistant password}
If a user wishes to enter his own key
number except the generated one by the system, then he/she can do so. To do
this, a user must follow a few instructions which are set by us. And we have named it
as Shoulder surfing \cite{b17} resistant password. The instructions are described as follows.
A user needs to add the decimal digits of the date of the current day in which
he/she is login to the system in such a way that it will form a single digit (0 to
9). Like if the current date is 27, then the sum of these two digits is (2+5=7). By
repeating this single digit 4 times (i,e 7777) it will form a 4-digit number. Then
user has to add this repeated 4-digit number with the original 4-digit key. If the
summation is greater than 10 then only the least significant digit will be taken.
As example if the original key is 1241 then after adding 7777 with it [Fig-2].
\begin{figure}[h]
\centering
\centerline{\includegraphics[width=6cm,height=6cm]{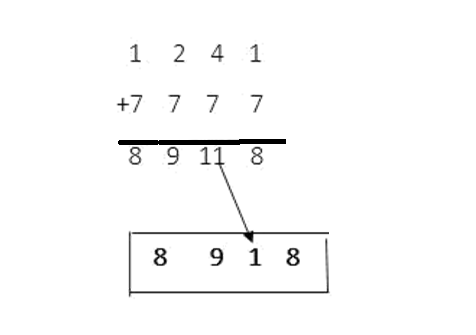}}
\caption{Generating Shoulder Surfing Resistant key} 
\label{fig2}
\end{figure}

Here the summation of 4+7=11 so the least significant digit of summation which is 1
will be taken only. Hence, the final key will be 8918. Then this will be the new
key number which can also be used for login purpose to avoid shoulder surfing.

System decodes the original key by subtracting 4-digit repeated number which
is formed from current date following the same procedure during encoding[Fig-3].
\begin{figure}[h]
\centering
\centerline{\includegraphics[width=4cm,height=3cm]{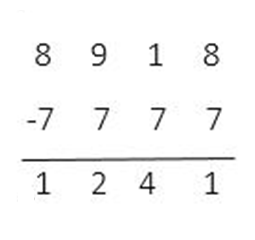}}
\caption{Extraction Of Original Key} 
\label{fig3}
\end{figure}
If any digit of the new shoulder surfing resistant key (minuend) is less than the
4-digit repeated number (subtrahend),the system will add 10 with the minuend
and then will subtract it. Here the 3rd digit 1 is less than 7 so system adds 10
with 1(1+10= 11) and 11-7= 4 is taken.
By doing this, system generates the original key and this key is cross checked
with the key saved in database.

\subsubsection{Architecture of the whole System }
Our system has two phases which interact between user and system. Mainly One is registration phase another is log in phase. The whole architecture is showing below with a Flow diagram. 

\begin{figure}[htbp]
\centering
\centerline{\includegraphics[width=9.5cm,height=18cm]{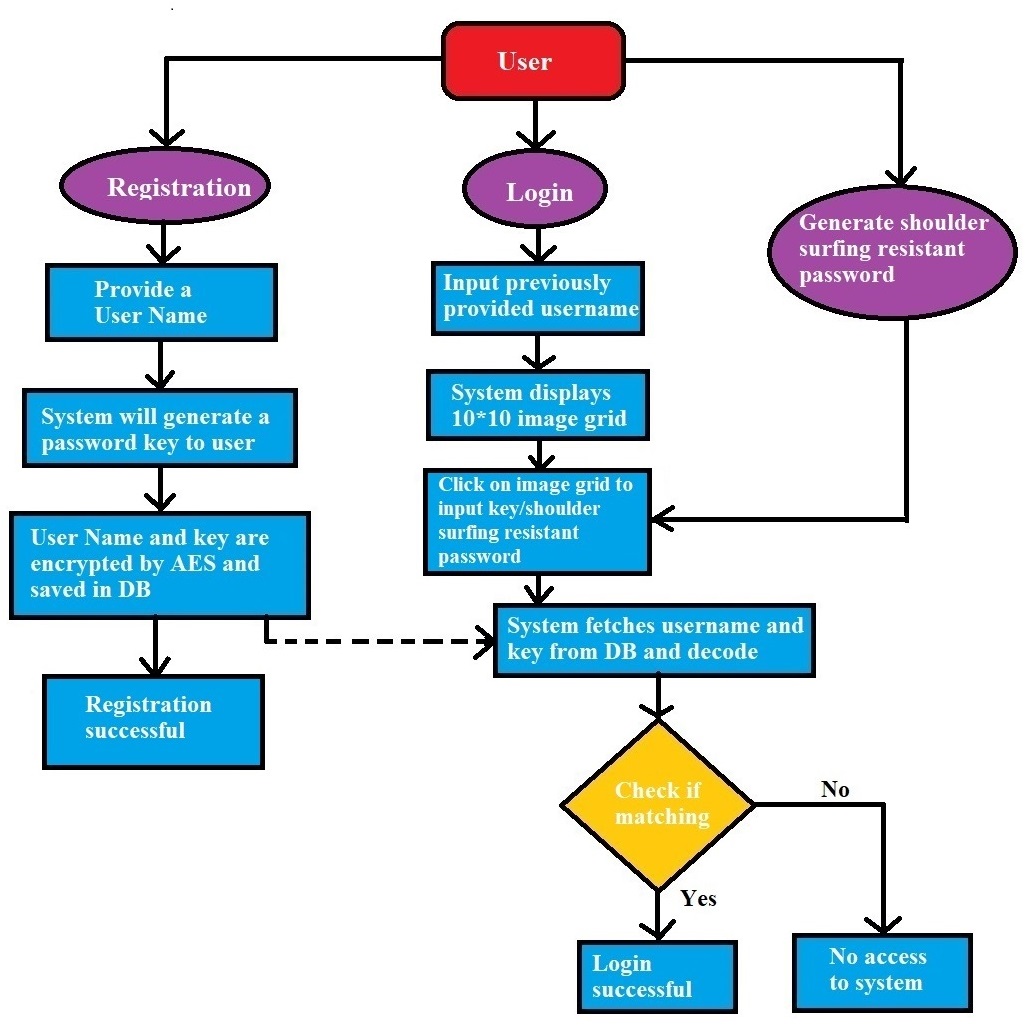}}
\caption{Flow Diagram of System Architecture} 
\label{fig4}
\end{figure}

\section{Advantage and Limitation}
\subsection{Advantage}
Our system will be able to prevent most of the chances of shoulder surfing which is secure for any system. It also subdues most of the drawbacks of textual password system.The Shoulder surfing resistant password mechanism is unique. The randomization speed of image grid is good enough which makes our system high performing and shoulder surfing resistant. Moreover, percentage of database hacking is almost low. In case of hacking occur, the information (key number and username) will be safe as these are encrypted and then saved in database. User memorization skill is not a big concern here. User just need to memorize the user name and key number and nothing else. So, all these features altogether provide huge advantage which makes our system efficient.
\subsection{Limitation}
User name is textual which is a limitation here. To make shoulder surfing registrant password, we are only using one factor (current date) to randomize the key. No recovery system is included in our system yet which is
a considerable limitation. Use of Static key for AES encryption makes our system a little insecure.

\section{Evaluation and analysis}
By analyzing our developed system performance, we have reached to the conclusion that
our system is quite high in performance. All the modules are designed in a user
friendly and efficient way. The image grid randomization speed is expectedly
good and every time shuffle properly. In short ,the system performance is according to our expectation.The below figure is showing the successful registration and login phase of our system.
\begin{figure}[htbp]
\centering
\centerline{\includegraphics[width=7cm,height=7cm]{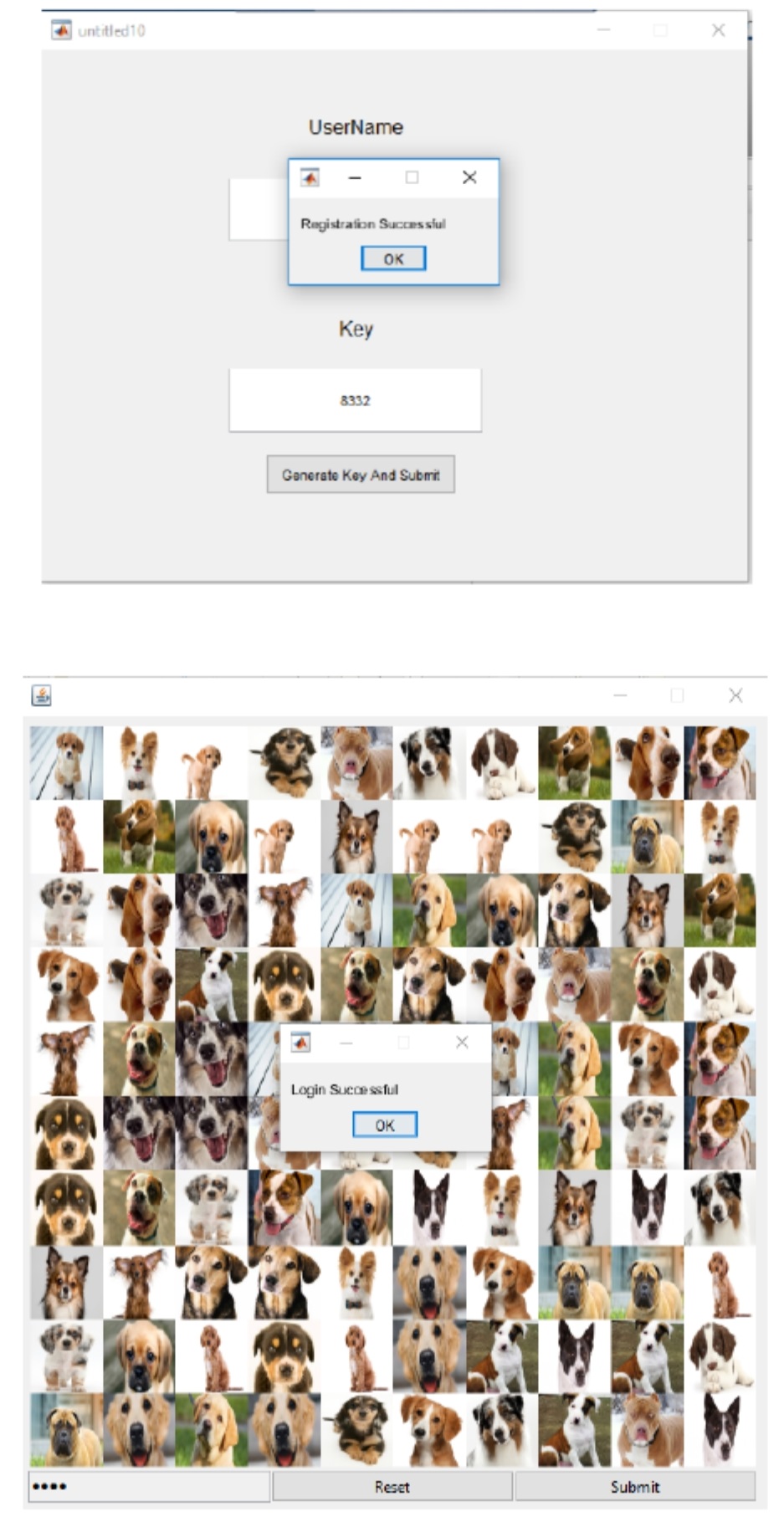}}
\caption{successful implementation of registration and login phase} 
\label{fig5}
\end{figure}
\section{Future Work}
Using stenography process any text can be hidden in a image. This image is called stego image. In future username will be hidden in image and will be provide to user for login  purpose. This will replace the criteria of providing user name for log-in. If a user wish to login, then the system will initially ask for that stego object from user in which the user name is hidden. A portable storage device like pen drive can be used to store that stego image.
\\
\\Initially we are encrypting the key number and user name using a static AES
key. Further we will use a randomize key which will enrich the security of the
system.
\\
\\In future we will add an extra feature in our system which will prevent brute force attack. That is by entering wrong key number 5 times the during login, system will block that user for almost half an hour. This way hackers attempt will go in vain if he tries brute force technique.
\section{Conclusion}
In this paper, we are proposing a new method of graphical authentication system and shoulder surfing resistant password mechanism which will enable us to overcome the shortcomings of password leakage. Moreover, the proposed system is free from ancient textual based password. It will be an efficient tool for protecting highly confidential data and information which will add a new dimension to the security purpose.

\ifCLASSOPTIONcaptionsoff
  \newpage
\fi

\end{document}